\documentclass{acm_proc_article-sp}

\usepackage{amsmath}
\usepackage{subfig}
\usepackage{pdfsync}
\usepackage{blindtext}
\usepackage[english]{babel}
\usepackage[utf8]{inputenc}
\usepackage{hyperref}
\usepackage{subfig}
\usepackage{float}
\usepackage{mathptmx}
\usepackage{xcolor}
\usepackage{graphicx}
\usepackage[final]{listings}
\usepackage{multirow}
\usepackage{footmisc}
\usepackage{fixltx2e}

\newcommand{\bq}{\begin{equation}}
\newcommand{\eq}{\end{equation}}

\newcommand{\flop}{\mbox{flop}}
\newcommand{\flops}{\mbox{flops}}

\newcommand{\FS}{\mbox{flop/s}}
\newcommand{\GBS}{\mbox{GB/s}}

\newcommand{\GHZ}{\mbox{GHz}}

\newcommand{\byte}{\mbox{byte}}

\newcommand{\cycle}{\mbox{cy}}
\newcommand{\cycles}{\mbox{cy}}
\newcommand{\cycl}{\mbox{cy/CL}}

\newcommand{\eos}{~.}

\newcommand{\construction}[1]{}%\marginpar{\includegraphics*[width=1cm]{Under_construction_icon-blue.eps}{#1}}}

\newcommand{\olsep}{\|}
\newcommand{\nolsep}{|}
\newcommand{\ecmspace}{\,}
\newcommand{\ecm}[6]{\mbox{$\left\{{#1}\ecmspace\olsep\ecmspace {#2}\ecmspace\nolsep\ecmspace {#3}\ecmspace\nolsep\ecmspace {#4}\ecmspace\nolsep\ecmspace {#5}\right\}\ecmspace{#6}$}}
\newcommand{\epsep}{\rceil}
\newcommand{\ecmp}[5]{\mbox{$\left\{{#1}\ecmspace\epsep\ecmspace {#2}\ecmspace\epsep\ecmspace {#3}\ecmspace\epsep\ecmspace {#4}\right\}\ecmspace{#5}$}}
\newcommand{\kerncraft}{Kerncraft}
\newcommand{\Kerncraft}{Kerncraft}

\newfont{\mycrnotice}{ptmr8t at 7pt}
\newfont{\myconfname}{ptmri8t at 7pt}
\let\crnotice\mycrnotice%
\let\confname\myconfname%

\begin{document}
\title{Automatic Loop Kernel Analysis and \\Performance Modeling With 
Kerncraft}

\numberofauthors{3}
\author{
\alignauthor
\makebox[0pt]{Julian Hammer\qquad Georg Hager\qquad Jan Eitzinger\qquad Gerhard Wellein}\\[1mm]
    \affaddr{\makebox[0pt]{Erlangen Regional Computing Center}}\\
    \affaddr{\makebox[0pt]{University of Erlangen-Nuremberg}}\\[1mm]
    \email{\makebox[0pt]{\{julian.hammer,georg.hager,jan.eitzinger,gerhard.wellein\}@fau.de}}}

\permission{Permission to make digital or hard copies of all or part of this work for personal or classroom use is granted without fee provided that copies are not made or distributed for profit or commercial advantage and that copies bear this notice and the full citation on the first page. Copyrights for components of this work owned by others than the author(s) must be honored. Abstracting with credit is permitted. To copy otherwise, or republish, to post on servers or to redistribute to lists, requires prior specific permission and/or a fee. Request permissions from Permissions@acm.org.}
\conferenceinfo{PMBS2015}{November 15 - 20 2015, Austin, TX, USA\\
{\mycrnotice{Copyright is held by the owner/author(s). Publication rights licensed to ACM.}}}
\copyrightetc{\the\acmcopyr.}
\crdata{ACM 978-1-4503-4009-0/15/11\ ...\$15.00\\
DOI: http://dx.doi.org/10.1145/2832087.2832092}

\maketitle
\abstract{Analytic performance models are essential for understanding 
the performance characteristics of loop kernels, which consume a major
part of CPU cycles in computational science. Starting from a validated
performance model one can infer the relevant hardware bottlenecks and
promising optimization opportunities. Unfortunately, analytic
performance modeling is often tedious even for experienced developers
since it requires in-depth knowledge about the hardware and how it
interacts with the software. We present the ``\kerncraft'' tool,
which eases the construction of analytic performance models for
streaming kernels and stencil loop nests. Starting from the loop source code, the
problem size, and a description of the underlying hardware, \kerncraft\
can ideally predict the single-core performance and scaling behavior
of loops on multicore processors using the Roof\/line or the
Execution-Cache-Memory (ECM) model. We describe the operating
principles of \kerncraft\ with its capabilities and limitations, and we
show how it may be used to quickly gain insights by accelerated
analytic modeling.}

\section{Introduction}\label{sec:intro}

This paper is concerned with analytic performance modeling on the CPU
core and chip level.  We define an \emph{analytic performance model}
as a simplified mathematical description of the interaction between
program code (``software'') and hardware, constructed with the goal in
mind to understand the dominating bottlenecks in program execution
performance.  The fundamental resources required for the execution of
any program are \emph{instructions} and \emph{data}; thus, in order to
predict the time it takes to execute a program one must have models
for how instructions are executed in the CPU core(s) and how data
travels through the memory hierarchy. It should be emphasized that the
more accurate a model the more specific to a CPU architecture it will
be. However, even simple analytic models can lead to surprising and
useful insights.

A simple, very optimistic model for instruction execution would assume
perfect out-of-order processing, perfect pipelining, and no
dependencies, with all required data residing in the L1
cache. Instructions would be distributed among the suitable pipelines,
and the pipeline that takes the longest time to execute its
instructions would determine the runtime of the code. We term this
model the \emph{throughput model} (TP). If dependencies along the
critical path are taken into account one arrives at a more pessimistic
model, the \emph{critical path model} (CP). In reality, and especially
for loop-based codes where some ``steady state'' execution may be
assumed, the actual runtime will be in between the TP and CP models as
long as no other bottlenecks apply, such as instruction cache misses,
instruction throughput limitations, or data transfers beyond the L1
cache.

The \emph{Roof{}line model} (see Sect.~\ref{sec:rl}) and the 
\emph{Exe\-cution-Cache-Mem\-ory (ECM) model} (see Sect.~\ref{sec:ecm}) are
two useful performance models for steady-state loop codes that build on
an instruction execution model but also take data transfers into
account. Both models may require considerable effort and experience to
construct in complicated cases (e.g., when the code is composed of
many loops, or when a loop has a complicated structure).
In this work we demonstrate a software tool, \kerncraft, that can automatically
construct these models from a C code formulation and a suitable hardware description without actually executing the code. 
We restrict ourselves to streaming and
stencil loop nests, where the data access patterns can be statically 
determined at compile time. It is not our intention to assess the usefulness or
accuracy of the models themselves, although some examples will be
given that highlight their capabilities.

Supporting tools are employed to determine parameters 
that are required as model input in the machine
description.  We use the LIKWID
tool suite \cite{10.1109/ICPPW.2010.38} for most of these tasks: The machine
topology, i.e., information about core and cache sharing, ccNUMA
structure, cache sizes, etc., is extracted from the output
of \verb.likwid-topology.. Achievable bandwidths to caches and main
memory are measured with
the \verb.likwid-bench. tool \cite{bench2012}, since it provides a
controlled and compiler-independent environment for building tailored
benchmark loops. 

Any analytic performance model must be checked for validity by comparing
its predictions with measurements on the target hardware.
The validation of predictions with measurements is an integral
part of the \kerncraft\ tool.

This paper is organized as follows: In
Sect.~\ref{sec:background} we briefly describe the components of the
performance models (in-core model, Roof\/line, and ECM) supported by
\kerncraft. Section~\ref{sec:testbed} introduces the hardware and
software used for all experiments. Details about the structure of the
\kerncraft\ tool and its concrete implementation are given in
Sect.~\ref{sec:kerncraft}.  In Sect.~\ref{sec:eval} we evaluate the
tool using streaming and stencil loop codes, and Sect.~\ref{sec:summary}
gives a summary and an outlook to future work.

The current version of \kerncraft\ is available for 
download at \linebreak\url{https://github.com/RRZE-HPC/kerncraft}.

\section{Background}\label{sec:background}

In this section we briefly describe the required components for
steady-state loop performance modeling: the in-core execution model,
the Roof\/line model, the ECM model, and approaches for model validation.

\subsection{In-core execution modeling}\label{sec:incore}

In simple cases the throughput and critical path modeling described above
can be done by hand. Listing~\ref{lst:ddot} shows the source code for
a scalar product in double precision.
\begin{lstlisting}[float=tb,label={lst:ddot},caption={Scalar product in double 
precision}]
double a[], b[], s=0.;
for(i=0; i<N; ++i)
  s += a[i] * b[i];
\end{lstlisting}
On a CPU with a three-stage ADD pipeline, the (scalar) naive code, i.e.,
without any unrolling and modulo variable expansion, would be limited
by the dependency on the summation variable \verb.s.. Hence, the core
would execute one \emph{floating-point operation} (\flop) every $3\,\cycles$,
which is also the CP 
prediction for one loop iteration. The TP prediction would be $2\,\cycles$ 
if the core can execute one LOAD, one ADD, and one MULT instruction
per cycle. With appropriate unrolling and modulo variable expansion
this limit can be achieved in practice if no other bottlenecks apply.

With more complicated code the pencil-and-paper analysis becomes
tedious and error-prone. One solution, albeit limited to Intel
architectures, is the \emph{Intel Architecture Code analyzer}
(IACA) \cite{iacaweb}, which provides TP and CP predictions for
assembly code sections. In \kerncraft\ we use the IACA TP output 
(and optionally the CP output)
as the in-core component in the ECM and Roof\/line models. IACA also provides
information about port utilization and points out a front-end 
bottleneck in case the code would need more concurrency than what
is supported by the instruction decoders. 

\subsection{Roof\/line model}\label{sec:rl}

The Roof\/line model is optimistic by design, i.e., it yields an
absolute lower execution time limit for a loop.
The ideas behind the model have been in use since the
1980s \cite{Kung:1986:MRB:17407.17362,Callahan88,hockney89}, but it
was popularized (and got its current name) by Williams et al.\ in 2009
\cite{roofline:2009}. It is based on the assumption that
performance is limited either by data transfers from a certain
memory hierarchy level or by the
computational (arithmetic) work, whichever takes longer.  This implies
that all data transfers perfectly overlap with computation. In
order to apply the model, the data volume from
and to each memory hierarchy level $k$, $\beta_k$, needs to be assessed 
and put in relation to the achievable peak bandwidth $B_k$ of that
level. Their ratio $T_k=\beta_k/B_k$ determines the data transfer
time for level $k$. To have a formulation that is, as far as possible, 
independent of the clock frequency setting, bandwidths are best
given in bytes per clock cycle. 

In a naive approach the time required for computation can be
calculated by dividing the work $\phi$ (usually \flops, but any other
well-defined metric will do) by the applicable computational peak
performance $P_\mathrm{peak}$ of the code at hand:
$T_\mathrm{core}=\phi/P_\mathrm{peak}$ (again, $P_\mathrm{peak}$ may be 
given in \flops/\cycle\ for a frequency-independent analysis).
However, an in-core execution model
as described in Sect.~\ref{sec:incore} is usually more accurate.
The final execution time prediction $T$ is then
$T_\mathrm{roof}=\max_k\left(T_\mathrm{core},T_k\right)$.
No substantial changes are required to adapt the model for multiple
cores; the achievable bandwidth $B_k(n)$ on $n$ cores in memory level $k$ must
of course be determined by measurements, and the data traffic and
code execution characteristics must take changes implied by
parallelization into account (e.g., if data is shared among 
multiple cores using the outer-level cache). 

\subsection{ECM model}\label{sec:ecm}

The Execution-Memory-Cache (ECM) model requires the same information
about the kernel code as the Roof\/line model. In contrast to
Roof\/line, it drops the assumption of a single bottleneck: Transfers
of data through the memory hierarchy are serialized across the
hierarchy levels of a core and therefore contribute to the reduction of the
total performance without overlapping with one another. The bandwidths
associated with each cache level are not taken from benchmarks, which
would also include contributions from all higher (closer to registers)
cache levels, but from published documentation by the vendors. The 
only measured input is the saturated maximum bandwidth of the 
memory interface. Since a cache line is the ``atomic'' data package in
the system, it is convenient to formulate the model with
a certain ``unit of work'' in mind, usually a number of iterations
that leads to a small integer number of cache line transfers (this 
pertains to the Roof\/line model as well). 

On a machine with three cache levels, there are five contributions to
the total single-core performance. We write them in a compact notation (see
\cite{stengel14} for details):
$$
\ecm{T_\mathrm{OL}}{T_\mathrm{nOL}}{T_\mathrm{L1L2}}{T_\mathrm{L2L3}}{T_\mathrm{L3MEM}}{}\eos
$$
$T_\mathrm{OL}$ and $T_\mathrm{nOL}$ are the overlapping and non-overlapping
times that come out of the in-core model (see Sect.~\ref{sec:kerncraft} for 
details on how these numbers are determined on a specific architecture). 
The non-over\-lapping contribution is serialized with the times
for the data transfers between adjacent memory levels $T_\mathrm{L1L2}$,
$T_\mathrm{L2L3}$, and $T_\mathrm{L3MEM}$, whereas $T_\mathrm{OL}$
overlaps with all data transfers. Note that the data transfer times
are not identical to the $T_k$ in the Roof\/line model, since they strictly
measure the time for getting data from one memory level to the next; 
the $T_k$, on the other hand, quantify the full transfer time from level $k$
into the registers.
A runtime prediction for a data
set in memory would be
$$
T_\mathrm{ECM,Mem} = \max\left(T_\mathrm{OL},T_\mathrm{nOL}+T_\mathrm{L1L2}+
T_\mathrm{L2L3}+T_\mathrm{L3Mem}\right)\eos
$$
For summarizing predictions for data in different memory levels we
use the following notation: 
$$
\ecmp{T_\mathrm{ECM,L1}}{T_\mathrm{ECM,L2}}{T_\mathrm{ECM,L3}}{T_\mathrm{ECM,Mem}}{}\eos
$$
In contrast to the Roof\/line model, the ECM model is not optimistic by
design, since it drops the single-bottleneck view. In practice, however,
it is rare that performance measurements exceed the model's predictions.

For multicore scaling, the model assumes perfect scalability until a
bandwidth bottleneck (usually the main memory bandwidth) is hit. It
thus predicts the number of cores where the loop performance ceases to
scale: $n_\mathrm s=T_\mathrm{ECM,Mem}/T_\mathrm{L3Mem}$. At that point, 
the ECM model prediction is identical to the bandwidth-based Roof\/line
prediction. See~\cite{stengel14} for more validations of the ECM model. 

\subsection{Model validation}

A performance model should be validated for correctness by running the
code on the target hardware. Different levels of validation are possible. 
In the simplest case, the measured runtime or performance is compared
with the prediction. Even if those numbers agree, it may still be the case
that several errors cancel out and the agreement is purely accidental. 
Performance counter measurements can be used to verify
quantities that come out of the model beyond the pure runtime, e.g., 
transferred data volume, memory bandwidth utilized, cache misses, etc.

In many cases it is particularly useful to vary problem and execution
parameters such as problem sizes, core counts, floating-point
precision, etc., in the model verification process, so that the 
model can be checked at many points in this configuration
space. This also leads to an improved understanding of the 
inherent bottlenecks by identifying typical performance patterns
\cite{proper12}.

\section{Testbed}\label{sec:testbed}

We choose two recent Intel CPU architectures for showing the modeling
results obtained with \kerncraft: Sandy Bridge and\linebreak Haswell. See
Table~\ref{tab:testbed} for a summary of documented hardware
properties. Note that measured inputs for the Roof\/line and ECM
models are not included for brevity. They are contained in the machine
description files distributed with the \kerncraft\
tool \cite{kerncraft-github}.

The Haswell system was configured to use ``Cluster on Die'' (CoD)
mode.  In CoD, one socket is effectively split into two ccNUMA domains
(with seven cores each in our case), with two memory controllers per
domain and separate L3 caches. The single-core ECM model uses the
saturated memory bandwidth of one ccNUMA domain for
$T_\mathrm{L3Mem}$. In addition, we have deactivated ``Uncore clock
frequency scaling,'' a power saving feature that reduces the clock
speed of the Uncore part of the chip (L3 cache and memory controllers)
when only one or two cores are active. All measurements were done with
the clock speed fixed to the base frequency (no ``Turbo Mode'') on
both systems. None of our benchmarks showed the documented
core slowdown feature (reduced base clock speed with highly efficient
vectorized code) on Haswell~\cite{hackenberg15}.

The software components used were the Intel compiler in version~15, 
LIKWID~4.0, Python~2.7, and IACA~2.1. 
\begin{table}
\begin{minipage}{\columnwidth}
    \centering
\begin{tabular}{|r|c|c|}
    \hline
    Microarchitecture   & SandyBridge-EP    & Haswell-EP \\
    \hline
    \hline
    Abbreviation        & SNB               & HSW \\
    Xeon model name     & E5-2680           & E5-2695 v3 \\
    \hline
    Clock (fixed)       & $2.7\,\GHZ$       & $2.3\,\GHZ$ \\
    Cores/Threads       & 8/16              & 2$\times 7$/28 \\
    \hline
    AVX throughput      & 1\,LD \& $\frac{1}{2}$\,ST & 2\,LD \& 1\,ST \\
    scalar throughput   & 2\,LD }{ 1\,LD \& 1\,ST & 2\,LD \& 1\,ST \\
    ADD throughput      & 1/cy              & 1/cy \\
    MUL throughput      & 1/cy              & 2/cy \\
    FMA3 throughput     & unsupported       & 2/cy \\
    \hline
    L1-L2 bandwidth     & $32\,\mbox{B/cy}$ & $64\,\mbox{B/cy}$\\%\footnote{although $64\,\mbox{B/cy}$ are given in documentation, we were unable to reproduce such results and utilized the better matching $32\,\mbox{B/cy}$.} \\
    L2-L3 bandwidth     & $32\,\mbox{B/cy}$ & $32\,\mbox{B/cy}$ \\
    L3-MEM bw. (peak)   & $51.2\,\GBS$      & $68.3\,\GBS$ \\
    \hline
\end{tabular}
\caption{Testbed hardware specifications and relevant microarchitecture features.
                 Note that the theoretical memory bandwidth is not used for
                 modeling.}
\label{tab:testbed}
\end{minipage}
\end{table}

\section{The kerncraft tool and supporting software}\label{sec:kerncraft}

\subsection{Overall architecture}\label{sec:arch}

\begin{figure}[tb]
    \includegraphics[width=\columnwidth]{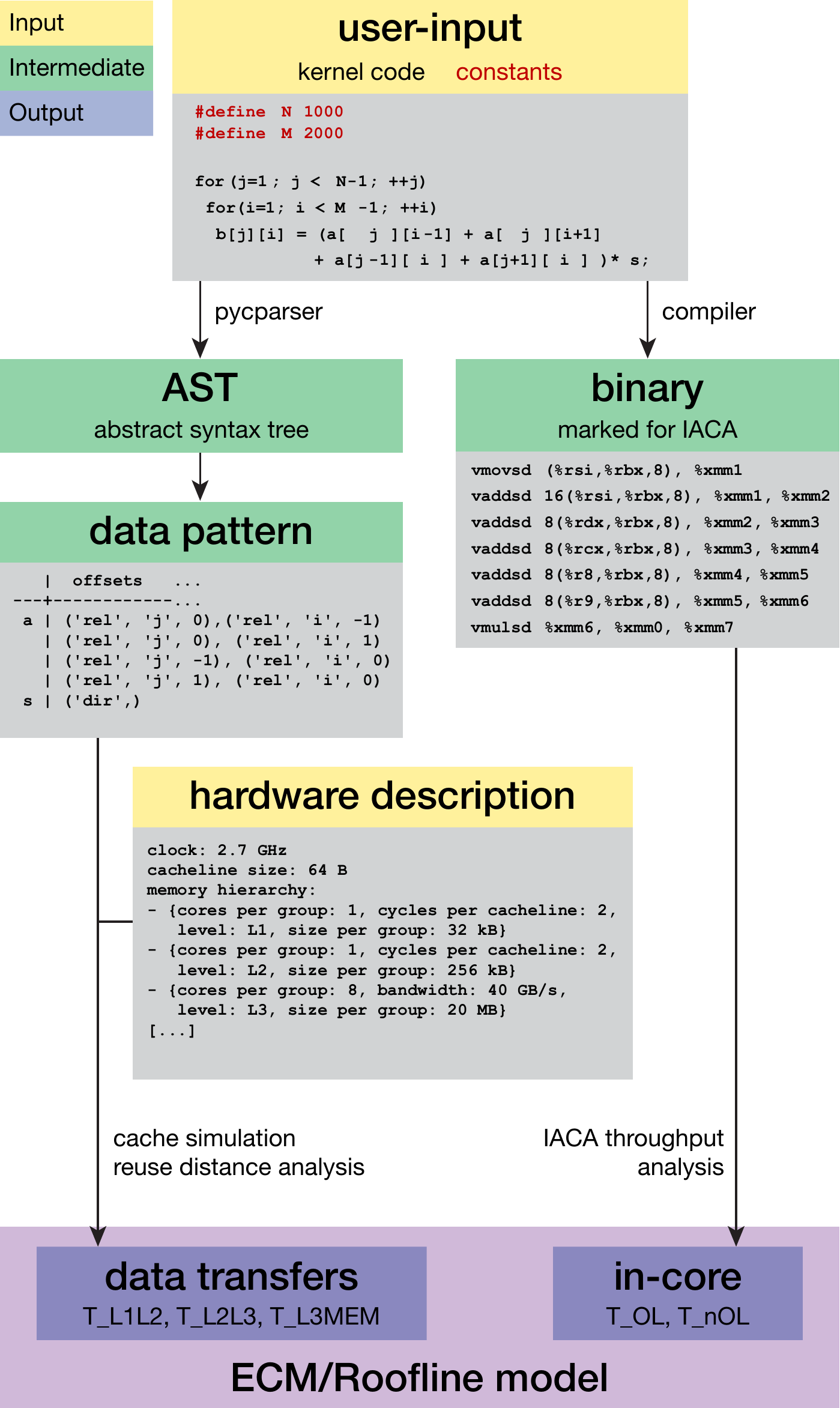}
    \caption{Overview of \kerncraft\ analysis}
    \label{fig:kerncraft-overview}
\end{figure}

\kerncraft\ \cite{kerncraft-github} combines static source code analysis
with static assembly analysis and cache access simulation to yield
performance predictions using the ECM and Roof\/line 
models for steady-state loop kernels. The kernel is specified
using the C language with some restrictions
(see Sect.~\ref{sec:code_input}). Static code
analysis yields information about the number of \flops, the data
structures, and the data volumes in the different memory hierarchy 
levels. The source is compiled and the
machine code is handed to IACA, which calculates an optimistic
throughput (TP) prediction for the loop body execution time.  In
addition, \kerncraft\ requires a hardware description file 
with architectural properties. 
With data transfers and
in-core predictions in place, all information is put together to 
yield Roof\/line and ECM predictions. For a general overview of the 
\kerncraft\ analysis steps see Fig.~\ref{fig:kerncraft-overview}.
Details about the different steps are given in the following sections.

The ECM and Roof\/line analysis can be executed on any machine with
the required software installed, since the code itself is not
actually run.  An optional ``Benchmark'' mode executes the compiled
kernel code to measure the performance rather than predicting it,
thereby validating the model(s).

In its current version the tool makes certain assumptions about the
hardware and software that go beyond those required for the Roof\/line
and ECM models: a perfect LRU cache replacement strategy and fully
associative, fully inclusive, and write-allocate/write-back
caches. These fit best to current Intel CPUs, but can be lifted to
adapt the framework to other architectures. Furthermore, the compiler
is assumed to not alter the data access characteristics of the
``as-is'' loop source code by, e.g., automatic blocking. Unless the
tool is integrated in a compiler, this restriction is hard to
circumvent.

\Kerncraft\ and its support scripts are implemented in 
Python 2.7 and are available on the Python Package Index (PyPI).

\subsection{Hardware description}

The hardware description file contains information about the
microarchitecture, the system architecture (e.g., number of cores and
sockets), and microbenchmark results. It can be created by hand
or with the help of the support
script \texttt{likwid\_auto\_bench.py}, which automatically creates a
formatted YAML \cite{YAML} file by collecting the required information
using \verb.likwid-topology. and \verb.likwid-bench.  on the machine
it is executed on. It performs microbenchmark runs with several
streaming kernels
in all memory
hierarchy levels and with all possible numbers of cores and threads.

In addition to the automatically gathered information, cache transfer
speeds, compiler flags, and the categorization of execution ports into
overlapping and non-overlapping, among others, need to be manually added
to the generated file.

Hardware information is required for all \kerncraft\ analyses, but it
can be gathered beforehand, on the targeted machine, and reused for
analysis runs on any other system. The hardware description
files for the architectures used in this paper are distributed
with the \kerncraft\ framework. Listing~\ref{lst:hw_desc} contains the
hardware description file for the Sandy Bridge machine described in Table~\ref{tab:testbed}, we left out the benchmark results and L2 and L3 cache
descriptions for brevity.

\begin{lstlisting}[float=t,
    caption=Hardware description for Intel Xeon E5-2680 Sandy Bridge with 2 Sockets,
    label={lst:hw_desc}]
clock: 2.7 GHz
cores per socket: 8
model type: Intel Core SandyBridge EP processor
model name: Intel Xeon CPU E5-2680 @ 2.70GHz
sockets: 2
threads per core: 2
cacheline size: 64 B
icc architecture flags: [-xAVX]
micro-architecture: SNB
FLOPs per cycle:
    SP: {total: 16, ADD: 8, MUL: 8}
    DP: {total: 8, ADD: 4, MUL: 4}
overlapping ports:
    ["0", "0DV", "1", "2", "3", "4", "5"]
non-overlapping ports: ["2D", "3D"]
memory hierarchy:
- {level: L1, cores per group: 1,
   cycles per cacheline transfer: 2,
   groups: 16, bandwidth: null,
   size per group: 32.00 kB,
   threads per group: 2}
# [... continues with description of L2 and L3]
- {level: MEM, cores per group: 8,
   cycles per cacheline transfer: null,
   groups: 1, bandwidth: null,
   size per group: null,
   threads per group: 16}
benchmarks:
  kernels:
    copy:
      FLOPs per iteration: 0
      read streams:
          {bytes: 8.00 B, streams: 1}
      read+write streams:
          {bytes: 0.00 B, streams: 0}
      write streams:
          {bytes: 8.00 B, streams: 1}
    # [... descriptions of daxpy, load, triad
    #      and update follow...]
  measurements:
# [... bandwidth measurements follow ...]
\end{lstlisting}

\subsection{Code input}\label{sec:code_input}

\begin{lstlisting}[float=tb,
    caption=2D 5-point Jacobi kernel code,
    label={lst:kernelcode}]
double a[M][N];
double b[M][N];
double s;

for(int j=1; j<M-1; ++j)
  for(int i=1; i<N-1; ++i)
    b[j][i] = (a[j][i-1] + a[j][i+1]
              +a[j-1][i] + a[j+1][i]) * s;

}}}
\end{lstlisting}

The source code that is subject to the analysis must be provided in a separate
file.  The required syntax is based on the ISO C99
standard~\cite{ISO:C99}. Constants (e.g., problem sizes)  can be passed  
on the command line. To simplify the analysis, there are currently some 
restrictions on the full C99 standard. For instance, array declarations 
may only use fixed sizes or constants, with an optional addition or 
subtraction of an integer (e.g., \texttt{double u[N][M+3][N-2][5]}, 
but not \texttt{double u[M*N]}), and array indices must use a loop index variable 
(with optional addition or subtraction), constants, or fixed integers.
A full list of limitations may be found in the software documentation. 
Listing~\ref{lst:kernelcode} shows a Jacobi kernel code with constants 
\texttt{M} and \texttt{N} defining the problem size in the two dimensions.

The pycparser~\cite{pycparser} Python library is used for parsing the
code. The resulting abstract syntax tree (AST) is then validated in
regard to the mentioned restrictions and statically analyzed. We
gather the following information: loop stack, data sources and
destinations, and floating point operations (\flop).
\begin{table}[tb]
    \centering
    \begin{tabular}{c|ccc}
    index variable & start & end & step size\\
    \hline
    \hline
    j & 1 & 499 & +1\\
    \hline
    i & 1 & 4999 & +1\\
    \hline
    \end{tabular}
    \caption{Loop stack of Listing~\ref{lst:kernelcode} with N=5000 and M=500}
    \label{tab:loopstack}
\end{table}
The loop stack contains information about all \verb.for. loops in the code:
their order, index variable name, start value, end value, and step
size. See Table~\ref{tab:loopstack} for the 
loop stack information gathered from Listing~\ref{lst:kernelcode}.
\begin{table}[tb]
    \centering
    \begin{tabular}{c|ll}
    variable & 1\textsuperscript{st} dimension & 2\textsuperscript{nd} 
dimension \\
    \hline
    \hline
    a & relative $j$ & relative $i-1$ \\
      & relative $j$ & relative $i+1$ \\
      & relative $j-1$ & relative $i$ \\
      & relative $j+1$ & relative $i$ \\
    \hline
    s & direct \\
    \hline
    \end{tabular}
    \caption{Data sources of Listing~\ref{lst:kernelcode} with N=5000 and M=500}
    \label{tab:datasources}
\end{table}
\begin{table}[bt]
    \centering
    \begin{tabular}{c|ll}
    variable & 1\textsuperscript{st} dimension & 2\textsuperscript{nd} 
dimension \\
    \hline
    \hline
    b & relative $j$ & relative $i$ \\
    \hline
    \end{tabular}
    \caption{Data destinations of Listing~\ref{lst:kernelcode} with N=5000 and M=500}
    \label{tab:datadestinations}
\end{table}
Data sources and destinations are retrieved from the statements in the
innermost \verb.for. loop. Any data access can be either direct (e.g., to a
scalar or an array with constant or fixed index) or relative with an
optional offset.  Multidimensional arrays can have mixed accesses; for
instance, \texttt{xy[0][j][i+1]} is a direct access on the first
dimension, relative on the second, and relative with an offset of +1
on the third. See
Tables~\ref{tab:datasources} and~\ref{tab:datadestinations} for the
data access analysis of Listing~\ref{lst:kernelcode}.

Finally the floating point operation
count of the inner loop body is extracted, for cases where a simple
in-core model is required (e.g, if IACA is unavailable). Since this is 
based on the plain source code, compiler
optimizations such as common subexpression elimination or compile-time
evaluation, as well as more intricate dependencies on the in-core
parallelism are ignored, and can lead to inaccuracies.

\subsection{In-core prediction}
\label{sec:in-core_prediction}

We use the Intel Architecture Code Analyzer (IACA) to statically
analyze the in-core performance of the kernel code.  It is freely
available and provides accurate throughput and critical path
predictions for code execution under the assumption that all code and
data comes from the L1 caches. Its predictions are based on documented
as well as unreleased information about Intel CPUs from Nehalem to
Haswell. The drawback is that older microarchitectures and
architectures by other vendors are not supported.

IACA operates on binaries and therefore requires the code to be
compiled. To do so, the kernel code is transformed by inserting it
into a main function and adding boilerplate code to handle declarations and
constants specified on the command line.  All array
declarations are replaced by pointer declarations with heap
allocation using \texttt{\_mm\_malloc} in order to have arrays
that are aligned to 32-\byte\ boundaries. The use of heap memory 
prevents problems with stack size constraints.
The loop stack is left
untouched, but the inner loop's statements are changed to reflect the
changed declarations, by replacing multi-dimensional indexes with
single-dimension index arithmetic. 
To prevent the compiler from
eliminating parts of the benchmark code by optimization, calls to external dummy
functions are inserted. The resulting code for the 2D-5pt
Jacobi kernel is shown in Listing~\ref{lst:bench_code}.
\begin{lstlisting}[float=bt,
    caption={Runnable code (shortened) generated by \Kerncraft
    for IACA analysis with dummy function calls.},
    label={lst:bench_code}]
#include <stdlib.h>
void dummy(double *, ... );
extern int var_false;

int main(int argc, char **argv) {
  const int N = atoi(argv[2]);
  const int M = atoi(argv[1]);
  
  double *a = _mm_malloc(
      (sizeof(double)) * (M * N), 32);
  for (int i=0; i<(M*N); ++i) a[i] = 0.23;
  // [...] allocate and init b[]
  double s = 0.23;
  if (var_false) dummy(a, b, &s);

  asm("nop"); // to limit search in assembly
  for (int j=1; j<(M-1); ++j)
    for (int i=1; i<(N-1); ++i)
      b[i+(j*N)] = (a[(i-1)+(j*N)] 
                  + a[(i+1)+(j*N)]
                  + a[i+((j-1)*N)] 
                  + a[i+((j+1)*N)]) * s;
  asm("nop"); // to limit search in assembly
}
\end{lstlisting}

To narrow down the throughput analysis to the relevant loop section,
IACA needs to find special markers in the machine code. These are inserted by
our tool, modifying the assembly produced by the compiler. 
There is an option to insert markers as in-line assembly in the 
C source, but this interferes with compiler optimizations. 
Finding the correct block in the assembly code is done by
selecting the block with the most vector instructions. The user can
always fall back to an interactive procedure by providing
a command line switch (\texttt{-\--asm-block=manual}) to \kerncraft.

The unrolling factor is extracted from the marked section by analyzing
the loop counter increments used at the end of the selected
block. This is needed to allow interpretation of the throughput
analysis results from IACA. For example, if a loop is four times unrolled
and IACA predicts a throughput of $32\,\cycles$ then one single loop
iteration and update step takes $8\,\cycles$.

IACA provides the number of cycles, per execution port, 
that one unrolled iteration (i.e., one assembly code loop body)
takes. In case of the 2D 5-point Jacobi kernel on a Sandy
Bridge microarchitecture, it takes $16\,\cycles$ in the data portion
of ports 2 and 3 (``2D'' and ``3D''). The maximum cycles of the data
portions in ports 2 and 3 makes up the non-overlapping contribution,
thus one might conclude that it would be 16\,\cycles. However, one 
unrolled iteration encompasses two ``work packages'' (cache line
lengths), so the applicable non-overlapping time is only
$T_\mathrm{nOL}=8\,\cycles$. The specific port numbers that are
responsible for non-overlapping execution are
defined in the hardware description file.

The maximum cycles over all other ports make up the overlapping
portion. It is $18\,\cycles$ with the code above on a Sandy Bridge
microarchitecture, thus $T_\mathrm{OL}=9\,\cycles$.

\subsection{Data traffic analysis}\label{sec:traffic}

\begin{figure*}
	\centering
    \includegraphics[width=0.95\textwidth]{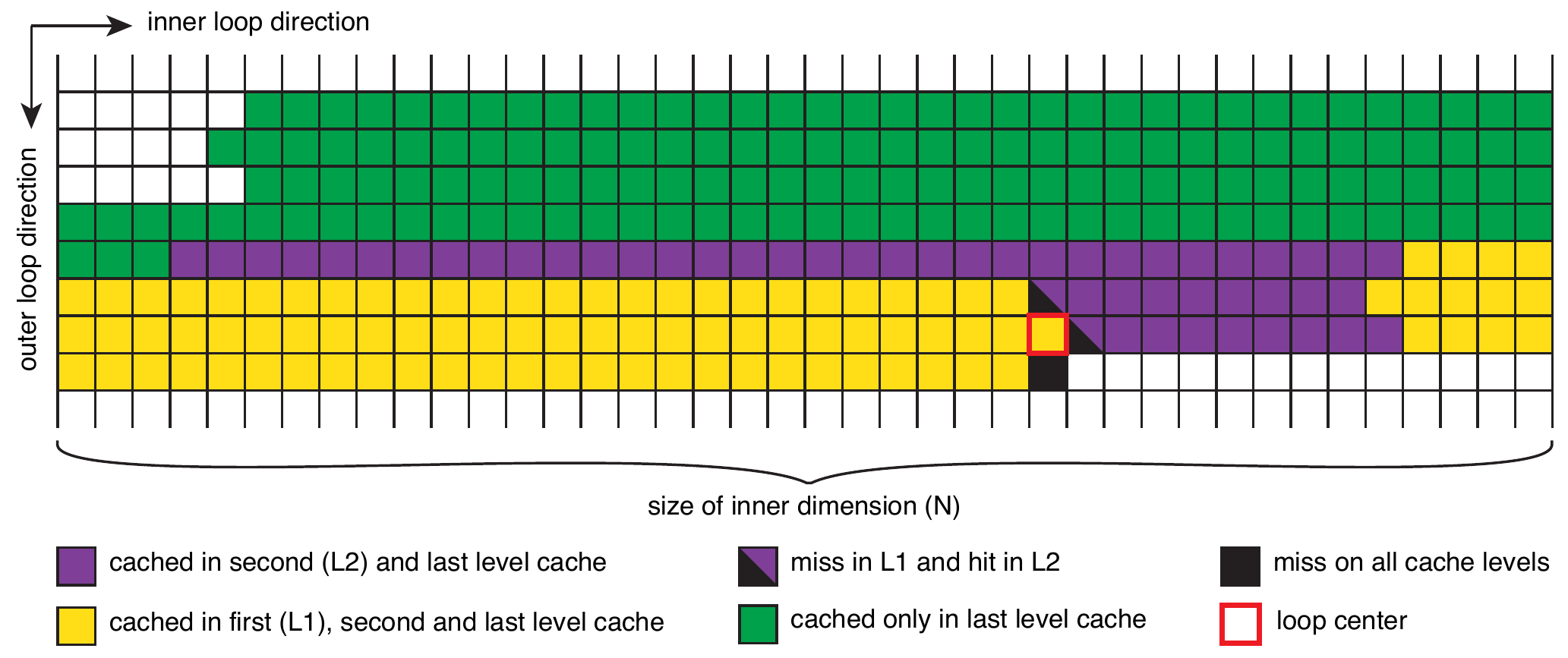}
    \caption{Cache usage prediction for the right-hand side array \texttt{a[]} 
    on the 2D-5pt Jacobi kernel, with $N=40$ on a hypothetical machine with 
    cache sizes that ensure the fulfillment of the layer condition 
    in L3 and in L2, but not in the L1 cache.}
    \label{fig:cacheusage}
\end{figure*}
The data traffic analysis is the central part of the \kerncraft\ tool,
enabling most of the insight when analyzing streaming and stencil loop
kernels. Its goal is to come up with predictions for the data volume
from and into all memory hierarchy levels, so that the cycle counts
required for the Roof\/line and ECM models can be determined. To this end
we have implemented a simple ``cache simulator'' (see Sect.~\ref{sec:arch}
above for the inherent assumptions about the hardware). In the following 
we explain its functionality using the 2D-5pt Jacobi example, since it
is a basic but nontrivial case where temporal reuse of data is crucial
for understanding performance properties.

Each cache level is inspected independently, although only misses in
lower levels (closer to registers) are passed to the next higher
level. We start off with all data requests from the loop iteration
$i$, $j$ in the code from Listing~\ref{lst:kernelcode} as 
shown in Tables~\ref{tab:datasources} and~\ref{tab:datadestinations}: \linebreak
\texttt{a[j-1][i]}, \texttt{a[j][i-1]}, \texttt{b[j][i]}, \texttt{a[j][i+1]}, \texttt{a[j+1][i]}. 
Unfortunately, this notation
does not offer any information about the actual location of the values
in memory, so we change to a
1D-notation: \texttt{a[(j-1)*N+i]}, \texttt{a[j*N+i-1]}, \texttt{b[j*N+i]}, \texttt{a[j*N+i+1]}, \texttt{a[(j+1)*N+i]}. The
mapping from 2D to 1D emphasizes the streaming nature of the data
accesses, but without $N$, $i$, and $j$, it is still not
sufficient. The inner problem size $N$ is specified by the user; we
choose $N=40$ here. The indices $i$ and $j$ are kept abstract, but we
assume a ``loop center'' at a relative offset of $0$ in either direction. 
If we plug this back into the 1D offsets, we are left 
with the following relative offsets: $-40$, $-1$, $0$, $+1$,
$+40$.

We now start adding single iterations backwards in the indices $i$ and
$j$ until the cache size is exceeded. After each addition of new
offsets, they are checked for overlaps with the original set of
accesses. If there is an overlap this is counted as a cache hit. Once
the cache is full, it is easy to see that all offsets from the
original set which were not turned into hits must be misses and thus
contribute to cache or memory traffic. See Figure~\ref{fig:cacheusage}
for an illustration of all three cache levels. Note that, if there are
many hits in the first level cache (L1), they are also hits in the
second and last level, but since only misses lead to utilization of
bandwidth, hits can be ignored. In our example we find one miss going
all the way to memory: It is the first access to a new cache line
in the outer loop direction (index $+40$, depicted in
Fig.~\ref{fig:cacheusage} as a black filled square).  
Also, there are two
misses on L1 to the right of the loop center and above it, at offsets
$-1$ and $+40$, but they are hits on L2. This situation can be expressed
in terms of the ``layer condition'' \cite{stengel14}: Three consecutive layers
of the grid fit into the L2 and L3 caches, but not into the L1 cache.
However, the L1 cache is certainly large enough to ensure that all accesses to
the left (index $-1$) will be L1 hits.

The actual communication between caches and memory is done in units of
cache lines. Therefore the number of cache lines
needs to be calculated from the offsets. We do not know any real
memory address, so we arbitrarily decide that the first cache-line
starts at offset $0$. Fig.~\ref{fig:cacheusage} contains all information 
to predict data amounts
originating from accesses to \texttt{a[]} between memory hierarchy
levels: the all-black cell is a miss throughout all cache levels. The
yellow cells are cached by L1 and therefore constitute
traffic between registers and L1, which is encapsulated in
$T_\mathrm{nOL}$ and given by IACA. The half-purple-half-black cells
are misses in L1, but hits in L2 and therefore generate traffic
between L1 and L2 and add to $T_\mathrm{L1-L2}$ with two cache lines.
Since everything was
already caught by higher cache levels, L3 has only the one original
miss to handle. 
In order to take care of write-allocates, all writes offsets are also
treated as reads.
All write offsets are added to an evict list and no caching is
tracked on this, meaning that all writes are immediately evicted as
long as they go to offsets in arrays. In our example, this leads to
an additional cache line transfer for each cache level.

\subsection{Model construction}

\Kerncraft\ currently supports six analysis modes:
\begin{description}

\item[\texttt{Roofline}] \hfill\\ Roof\/line with a simple in-core model based
                         on arithmetic peak performance and L1 cache as
                         an additional bandwidth bottleneck; does not require compiler
                         or IACA

\item[\texttt{RooflineIACA}] \hfill\\ Roof\/line with IACA for in-core modeling

\item[\texttt{ECM}] \hfill\\ Full ECM model with IACA for in-core modeling

\item[\texttt{ECMData}] \hfill\\ Data transfer portion of ECM only; does not require compiler
                        or IACA

\item[\texttt{ECMCPU}] \hfill\\ In-core (IACA-based) model only

\item[\texttt{Benchmark}] \hfill\\ Benchmark run for model validation

\end{description}
In the following we describe how the models in the first five 
modes are put together.

\subsubsection{Roof{}line}

The Roof\/line model can be constructed in two different ways, which are
distinguished by the in-core modeling: 
either by using IACA (\texttt{RooflineIACA} mode) or by calculating the
theoretical (MULT+ADD) arithmetic peak performance of the CPU 
(\texttt{Roofline} mode). 
If the theoretical peak is used, the L1 cache needs to
be considered as another cache level and potential bandwidth bottleneck 
as described below. The peak
performance is usually an over-optimistic estimate, because it assumes the
perfect mix of operations, full SIMD vectorization, etc., for the given 
CPU. A more realistic model
will replace this in future work.

Each memory level
(except the L1 cache in \texttt{RooflineIACA} mode) 
is considered a potential data transfer bottleneck.  As described
in Sect.~\ref{sec:rl}, cycle counts for the calculated data volume
are determined based on microbenchmark results in the hardware 
description, giving a minimum execution time for every memory
hierarchy level. Several microbenchmarks are available to 
provide a ``closest match'' to the actual loop code: e.g., if
one read stream, one write stream, and one write-allocate stream
hit a certain memory level, the measured bandwidth of an array
copy benchmark in that level is used as the bandwidth baseline.

Among all lower execution time bounds for all memory levels and for
the in-core execution, the largest is used as the Roof\/line model
prediction (single-bottleneck view). Apart from the prediction in
cycles per cache line, the tool also calculates the arithmetic
intensity for the memory level that was found to be the
bottleneck. For producing the output the cycle counts are by default
converted to performance numbers in \FS. This can be changed
via a command line option (\texttt{-\--unit cy/CL}). In general,
the units \texttt{cy/CL}, \texttt{It/s}, and \texttt{FLOP/s} 
can be selected. 

\subsubsection{ECM}

The ECM analysis mode is split into two sub-modes: \texttt{ECMData} for the
data access and cache analysis, and \texttt{ECMCPU} for the in-core
analysis performed by IACA. The model \texttt{ECM} combines both
submodels and gives an overall prediction. The reason for the split is
that IACA may not be available on all systems, and it does not support
microarchitectures other than Intel's. In such cases one may still use
the \texttt{ECMData} model to get the data traffic analysis.

Using the in-core prediction and the transfer times between adjacent
cache levels, the ECM model is put together and printed using the
syntax as described in Sect.~\ref{sec:ecm}. By default the report is
in \cycl, but the standard options (see the previous section) are
available to change this. The tool also reports the number of cores at
which the performance is expected to saturate.

Examples for the application and the output of the tool can be found in 
Sect.~\ref{sec:eval} below.

\subsection{Model validation (Benchmark mode)}

The \texttt{Benchmark} mode of \kerncraft\ does not
predict performance but measures it. Therefore this mode must 
be executed on the same machine as described in the hardware
description file. A matching hardware description is necessary to
determine the appropriate compiler flags, cache line sizes, and clock
speed. For the measurement results to be generated correctly, the user
must fix the CPU frequency to the clock speed defined in the machine
file.

In order to allow execution of the kernel code, it is transformed in a
similar manner as for IACA, described in
Section~\ref{sec:in-core_prediction}, but with one
extra \verb.for. loop inserted that wraps all kernel loops. This is to
prolong the total execution time for improved accuracy with small
problem sizes. Also, calls to the LIKWID marker application
programming interface are inserted before and after the
outer \verb.for. loop for a precise measurement of the loop section
only, without overhead from allocating and initializing memory. After
successful transformation and compilation, the binary is executed
with \texttt{likwid-perfctr}, LIKWID's performance counter and pinning
command. Proper thread-core affinity is therefore ensured.  The marker
calls can also be used for advanced validation using data volume,
cycles per instruction, and work-related metrics, but this feature is
not leveraged in the current version of \kerncraft.  The runtime is
used to calculate the number of cycles per work unit.

\section{Evaluation}\label{sec:eval}

Here we evaluate the correctness of predictions derived by \kerncraft\
from kernel codes and hardware descriptions. It is out of the scope of
this work to evaluate the underlying performance models; this has been
discussed elsewhere 
\cite{datta09,roofline:2009,CPE:CPE3180,stengel14,HFEHW15,hofmann2015}. We will,
however, compare predictions by \kerncraft\ to predictions
derived by manual analysis in previously published papers 
(see Table~\ref{tab:evaluation}) and point 
out relevant differences and peculiarities. We only consider
predictions for serial code on in-memory data sets here, although both models
are certainly capable of providing in-cache and multi-threaded 
results as well (supported by the \verb.--cores. option).
The same binary code has been produced for the Sandy
Bridge and the Haswell CPU, but IACA was run with the appropriate
architectural setting.
\begin{table*}
\begin{minipage}{\textwidth}
    \centering
\begin{tabular}{ccc|ccccc|cc}
       &      &       & \multicolumn{5}{|c|}{Kerncraft}               & \multicolumn{2}{|c}{Reference} \\
       &      &       & Const. & ECM Model & ECM\footnote{in-memory prediction\label{fn:mem}}  & Roofline\footref{fn:mem} & Bench. & ECM Model & Roofline\footref{fn:mem} \\
Kernel & Code & Arch. & $N$    & (\cycl)   & (\cycl)                           & (\cycl)                  &  (\cycl) & (\cycl) & (\cycl)\\
\hline
\hline
\multirow{2}{*}{2D-5pt} & \multirow{2}{*}{List.~\ref{lst:kernelcode}} & SNB & \multirow{2}{*}{$6000$} & \ecm{9.5}{8}{10}{6}{12.7}{} & $36.7$ & $29.8$ & $36.4$ & \ecm{6}{8}{10}{10}{13}{} \cite{stengel14} & n/a \\
 & & HSW & & \ecm{9.4}{8}{5}{6}{16.7}{} & $35.7$ & $26.6$ & $30.0$ & n/a & n/a \\
\hline
\multirow{2}{*}{UXX} & \multirow{2}{*}{List.~\ref{lst:uxx_code}} & SNB & \multirow{2}{*}{$150$} & \ecm{84}{32.5}{20}{20}{26.3}{} & $98.8$ & $84.0$ & $112.5$ & \ecm{84}{38}{20}{20}{26}{} \cite{stengel14} & n/a \\
 & & HSW & & \ecm{56}{27.5}{10}{20}{31.6}{} & $89.1$ & $61.7$ & $86.9$ & n/a & n/a \\
\hline
\multirow{2}{*}{long-range} & \multirow{2}{*}{List.~\ref{lst:long-range_code}} & SNB & \multirow{2}{*}{$100$} & \ecm{57}{53}{24}{24}{17.0}{} & $118.0$ & $65.9$ & $134.2$ & \ecm{68}{64}{24}{24}{17}{} \cite{stengel14} & n/a \\
 & & HSW & & \ecm{57}{47.5}{12}{24}{22.3}{} & $105.8$ & $63.6$ & $104.5$ & n/a  & n/a \\
\hline
\multirow{2}{*}{Kahan-dot} & \multirow{2}{*}{List.~\ref{lst:kahan-dot_code}} & SNB &  \multirow{2}{*}{}  & \ecm{96}{8}{4}{4}{7.8}{} & $96.0$ & $96.0$ & $101.1$ & \ecm{32}{8}{4}{4}{7.9}{}\footnote{without empirical penalties} \cite{HFEHW15} & n/a \\
 & & HSW & & \ecm{96}{8}{2}{4}{9.1}{} & $96.0$ & $96.0$ & $98.0$ & n/a & n/a \\
\hline
Schönauer & \multirow{2}{*}{List.~\ref{lst:triad_code}} & SNB & \multirow{2}{*}{} & \ecm{4}{6}{10}{10}{21.9}{} & $47.9$ & $54.3$ & $58.8$ & \ecm{4}{6}{10}{10}{24}{} \cite{CPE:CPE3180} & n/a \\
Triad & & HSW & & \ecm{4}{3}{5}{10}{26.3}{} & $44.3$ & $46.4$ & $48.3$ & n/a & n/a \\%\ecm{1}{4}{6}{10}{26.5}{} \\%\footnote{single precision floating-point arithmetic} \cite{hofmann2015} & n/a \\
\hline
\end{tabular}
\caption{Single-thread \kerncraft\ predictions for five benchmark kernels in comparison to 
                 reference predictions 
                 from previous publications.}
\label{tab:evaluation}
\end{minipage}
\end{table*}

\subsection{Stencil Codes}

We choose three interesting ``corner case'' stencils that have been
comprehensively analyzed in \cite{stengel14} using the ECM
model on a Sandy Bridge base microarchitecture: 
2D-5pt Jacobi, UXX, and a fourth-order long-range stencil. 

\subsubsection{2D-5pt Jacobi}

\begin{lstlisting}[float=tb,
    caption={\Kerncraft\ output for the analysis of the 2D-5pt Jacobi kernel on the Sandy Bridge microarchitecture (node name ``phinally'') with $N=6000$ and $M=6000$ (output adapted for brevity)},
    label={lst:jacobi_analysis},
    numbers=left,stepnumber=2]
$ kerncraft %-p ECM% --cores 1 \
            -m machine-files/phinally.yaml \
            kernels/2d-5pt.c \
            -D N 6000 -D M 6000
            
saturating at 3 cores

$ kerncraft %-p RooflineIACA %\
            %-{}-unit cy/CL% --cores 1 \
            -m machine-files/phinally.yaml \
            kernels/2d-5pt.c \
            -D N 6000 -D M 6000 -v

Bottlnecks:
 level | ar.int. | perfor. | bandw. | bw kernel
       |  FLOP/b |   cy/CL |   GB/s | 
-------+---------+---------+--------+----------
   CPU |         |     9.5 |        |
 L1-L2 |    0.1  |    16.9 |   51.2 | triad
 L2-L3 |    0.1  |    27.5 |   31.5 | triad
L3-MEM |    0.17 |    29.8 |   17.4 | copy

Cache or mem bound
           (bw with from copy benchmark)
Arithmetic Intensity: %0.17 FLOP/b%
\end{lstlisting}

In Listing~\ref{lst:jacobi_analysis} we show the \kerncraft\ command
line arguments and generated output for the 2D 5-point Jacobi stencil
(Listing~\ref{lst:kernelcode}) for Sandy Bridge. In line
number~\ref{l:ecm} we see the ECM model and line number~\ref{l:ecmp}
shows the ECM prediction notation. The results in
Table~\ref{tab:evaluation} have a difference of $3\,\cycles$ in
$T_\mathrm{OL}$ on Sandy Bridge compared to the analysis
in \cite{stengel14}.  This is due to the compiler using half-wide
(16-\byte) load instructions in our case.  The half-wide loads take the
same number of cycles as full-wide loads, but require two addresses to
be generated. In combination with an additional address required for
full-width store operations, this leads to an average of $9\,\cycles$
of address generation per cache-line. The same happens on Haswell,
although it has one additional address generation unit. However,
this unit can only be utilized with simple addressing modes, and
the compiler does not leverage this feature. On Haswell
the ECM model is outperformed by the measurement by almost 20\%. 
This is very unusual, although the model is not strictly optimistic
like Roof\/line, and will be investigated.

The Roof\/line model is much more optimistic than the ECM model for this
code, since the memory bottleneck does not dominate the single-core
execution time: The layer condition can only be satisfied in the L2
cache for the chosen inner problem size ($N=6000$). 
If spatial blocking for the L1 cache is performed (or if the
inner loop size is short enough), Roof\/line becomes more accurate as
shown in \cite{stengel14}.

\subsubsection{UXX (double precision)}

\begin{lstlisting}[float=tb,
    caption={Kernel code for the UXX stencil in double precision},
    label={lst:uxx_code},]
double u1[M][N][N], d1[M][N][N], xx[M][N][N];
double xy[M][N][N], xz[M][N][N], c1,c2,d,dth;

for(int k=2; k<M-2; k++) {
  for(int j=2; j<N-2; j++) {
    for(int i=2; i<N-2; i++) {
      d = (d1[k-1][j][i] + d1[k-1][j-1][i]
         + d1[k][j][i]  + d1[k][j-1][i])*0.25;
      u1[k][j][i] = u1[k][j][i] + (%dth/d%)
      * ( c1*(xx[k][j][i]   - xx[k][j][i-1])
        + c2*(xx[k][j][i+1] - xx[k][j][i-2])
        + c1*(xy[k][j+1][i] - xy[k][j-1][i])
        + c2*(xy[k][j+1][i] - xy[k][j-2][i])
        + c1*(xz[k][j][i]   - xz[k-1][j][i])
        + c2*(xz[k+1][j][i] - xz[k-2][j][i]));
}}}
\end{lstlisting}
The UXX stencil (Listing~\ref{lst:uxx_code}) has a long-latency divide
operation in the inner loop that cannot be easily avoided and that
dominates the overlapping time. This explains the large difference
between Sandy Bridge and Haswell in terms of $T_\mathrm{OL}$.
The discrepancy in $T_\mathrm{nOL}$ compared to the reference result
is due to the more current compiler, but there is also a deviation
between Haswell and Sandy Bridge: The compiler generates a mixture
of full-wide (32-\byte) and half-wide (16-\byte) load instructions.
Since both architectures can execute two half-wide loads per cycle,
but only Haswell can do two full-wide loads per cycle, Haswell
has a slight advantage in load throughput and, consequently,
in $T_\mathrm{nOL}$.

\subsubsection{Long-range}

\begin{lstlisting}[float=tb,
    caption={Kernel code for the long-range stencil},
    label={lst:long-range_code},]
double U[M][N][N], V[M][N][N], ROC[M][N][N];
double c0, c1, c2, c3, c4, lap;

for(int k=4; k < M-4; k++) {
  for(int j=4; j < N-4; j++) {
    for(int i=4; i < N-4; i++) {
      lap = c0*V[k][j][i]
          + c1*(V[k][j][i+1] + V[k][j][i-1])
          + c1*(V[k][j+1][i] + V[k][j-1][i])
          + c1*(V[k+1][j][i] + V[k-1][j][i])
          + c2*(V[k][j][i+2] + V[k][j][i-2])
          + c2*(V[k][j+2][i] + V[k][j-2][i])
          + c2*(V[k+2][j][i] + V[k-2][j][i])
          + c3*(V[k][j][i+3] + V[k][j][i-3])
          + c3*(V[k][j+3][i] + V[k][j-3][i])
          + c3*(V[k+3][j][i] + V[k-3][j][i])
          + c4*(V[k][j][i+4] + V[k][j][i-4])
          + c4*(V[k][j+4][i] + V[k][j-4][i])
          + c4*(V[k+4][j][i] + V[k-4][j][i]);
      U[k][j][i] = 2.f*V[k][j][i] - U[k][j][i]
                 + ROC[k][j][i] * lap; }}}
\end{lstlisting}
The fourth-order long-range stencil
(Listing~\ref{lst:long-range_code}) shows a very low fraction of
execution time in $T_\mathrm{L3Mem}$ due to a massive amount of data
(layers) that have to be supplied by the caches. The results in
Table~\ref{tab:evaluation} exhibit a slight deviation in $T_\mathrm{OL}$ of
$16\,\cycles$ and in $T_\mathrm{nOL}$ of $9\,\cycles$ on Sandy
Bridge compared to published results. 
The lower cycle predictions in our analysis can be attributed to better
compiler-generated code (fewer register spills). Again, Haswell
has a slight advantage over Sandy Bridge in terms of $T_\mathrm{nOL}$
cycles due to the compiler partially producing full-wide load
instructions. The Roof\/line prediction is massively
over-optimistic in all cases since it assumes a core-bound
situation, whereas in fact the execution time is spread almost
evenly between in-core and data transfer contributions.

The long-range stencil constitutes an instructive example for
demonstrating the complicated situations that can arise
with stencil codes on architectures with multiple caches. 
\begin{figure}[tb]
\includegraphics[width=\linewidth]{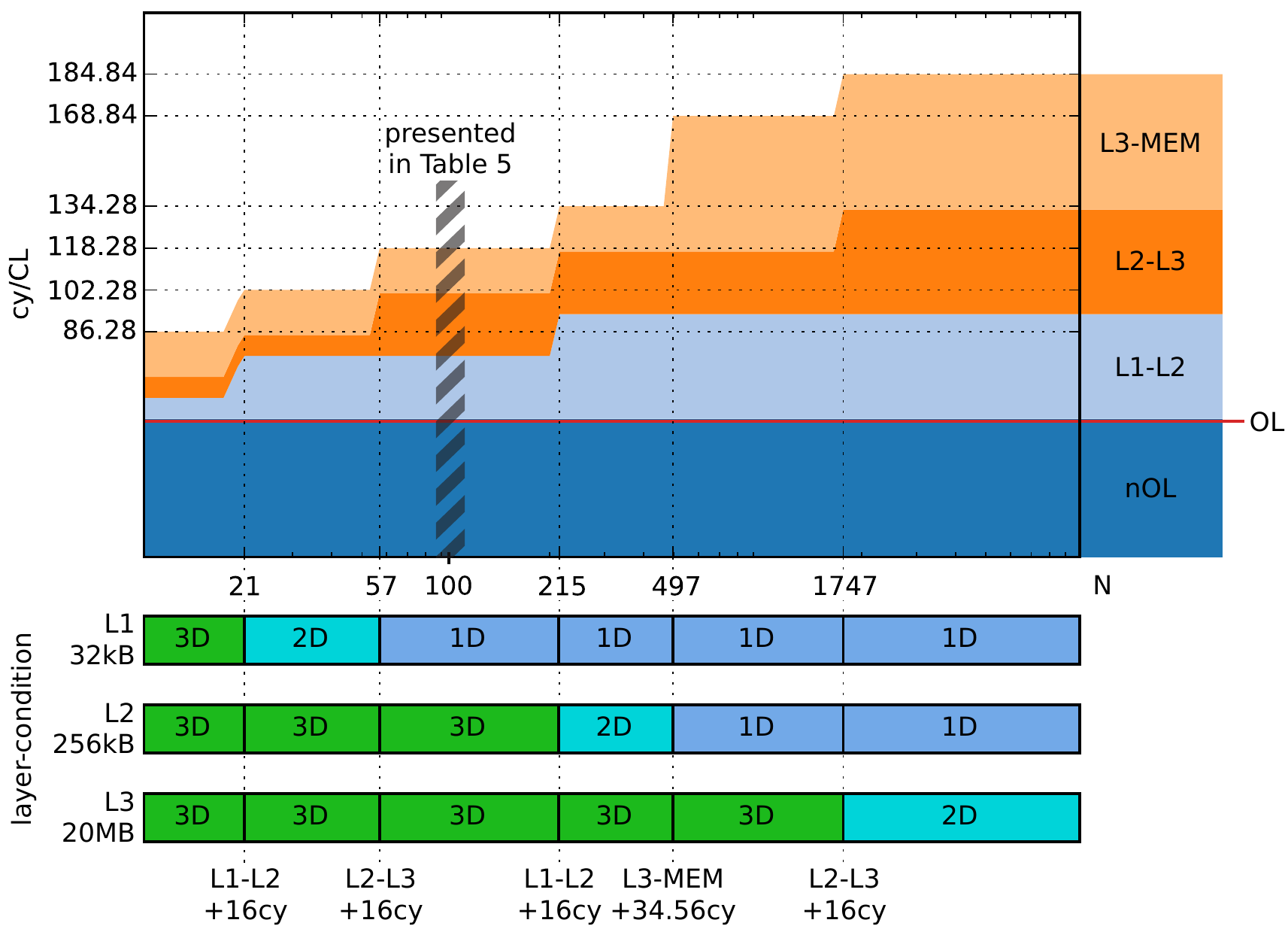}
\caption{\label{fig:3dlr-pred}Top: Single-core ECM model predictions 
for the 3D long-range stencil (Listing~\ref{lst:long-range_code}) with
increasing inner (and middle) dimension $N$ and a memory-bound data
set. Bottom: Visualization of fulfilled layer conditions in the three
cache levels. A 3D/2D/1D layer condition pertains to loop index $k$, $j$, 
and $i$, respectively.}
\end{figure}
In Fig.~\ref{fig:3dlr-pred} we show graphically the contributions to
the ECM model in cycles per cache line versus the inner and middle
loop size (both set to $N$). The entry in Table~\ref{tab:evaluation}
uses $N=100$ (marked in the figure). Varying $N$ leads to layer
conditions being fulfilled and violated in different cache levels;
overall, six situations can be distinguished over the $N$ range
considered, shown at the bottom of the figure: for instance, around
$N\approx 1000$ the layer condition in the outer ($k$) dimension is
met in the L3 cache alone, whereas the L2 and L1 caches are
too small to even hold a sufficient number of rows to meet
the layer condition in the middle ($j$) dimension.
\begin{figure}[tb]
\includegraphics[width=\linewidth]{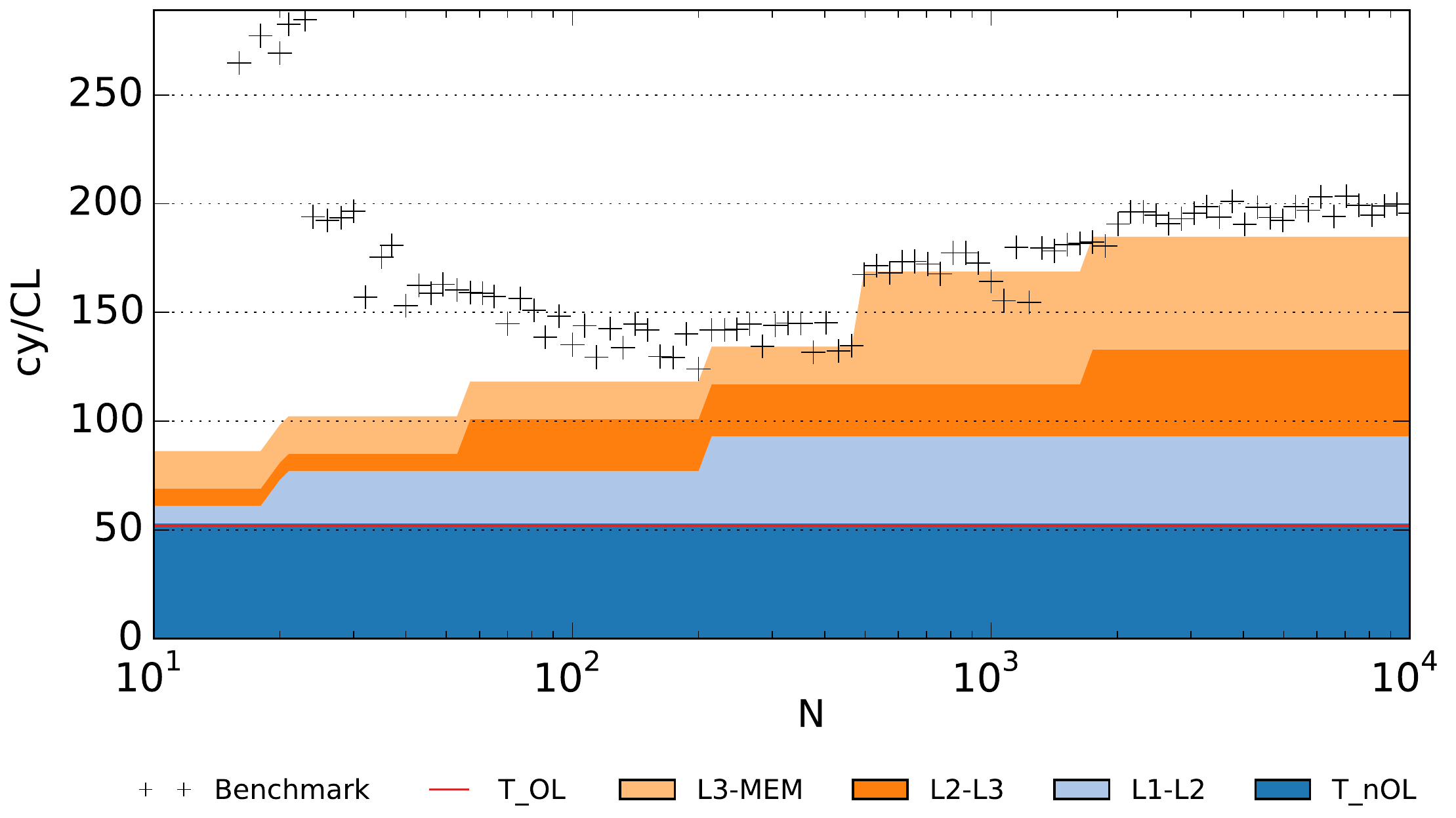}
\caption{\label{fig:3dlr-val}Validation of the single-core ECM 
prediction for the 3D long-range stencil (as presented in Fig.~\ref{fig:3dlr-pred}), with measurements (crosses) on SNB.}
\end{figure}
Figure~\ref{fig:3dlr-val} shows the predictions together with
measured data (crosses). The general behavior with increasing $N$
is well tracked by the model starting at $N\gtrsim 200$, but 
there are considerable deviations for smaller $N$. This is
expected since the steady-state assumption cannot be upheld
with short inner loops, and boundary effects become non-negligible.

\subsection{Streaming Kernels}

\Kerncraft\ does not only support the 
analysis of stencil codes, but also accepts typical streaming kernels
found in numerical vector arithmetic and memory benchmarks. Here we
will have a look at the Kahan-compensated double-precision dot product
and the Sch\"onauer Triad. These have been analyzed thoroughly 
in \cite{HFEHW15} and \cite{CPE:CPE3180}, respectively.

\newpage
\begin{lstlisting}[
    caption={Kernel code for Kahan-ddot},
    label={lst:kahan-dot_code},]
double a[N], b[N], c;
double sum, prod, t, y;

for(int i=0; i<N; ++i) {
    prod = a[i] * b[i]; y = prod - c;
    t = sum + y; c = (t - sum) - y; sum = t;
}
\end{lstlisting}

\subsubsection{Kahan-ddot}
The code is shown in Listing~\ref{lst:kahan-dot_code}. It provides a
scalar product of two arrays \verb.a[]. and \verb.b[]., correcting
for round-off errors due to the finite-precision floating-point number
representation~\cite{Kahan:1965}. As was shown in \cite{HFEHW15},
current compilers fail in generating efficient (or correct) machine
code from the C source. In our case the compiler could not use SIMD
vectorization due to the presence of a loop-carried dependency, but it
produced correct scalar code without further unrolling. As a
consequence the runtime is dominated by the loop-carried dependency
and IACA reports $T_\mathrm{OL}=96\,\cycl$ even in throughput mode.
Here we find a case where the ECM and Roof\/line predictions
coincide due to the strong dominance of $T_\mathrm{OL}$. 

The result for $T_\mathrm{nOL}$ on Sandy Bridge in
Table~\ref{tab:evaluation} differs from the reference result, since
the latter was produced with a scalar but otherwise optimal version of
the code that used modulo unrolling to hide the inter-iteration stalls.
Note also that \cite{HFEHW15} uses latency penalties to make the 
ECM model work better in memory. The \kerncraft\ tool has this
capability (in fact, the penalty cycles are part of the machine 
files), but it is deactivated by default.

\subsubsection{Sch\"onauer Triad}
The Sch\"onauer Triad (Listing~\ref{lst:triad_code}) is a simple streaming kernel
which is designed to be limited by data transfers
on all processor architectures even if the data
resides in the innermost cache.

The deviation of $T_\mathrm{L3-MEM}$ compared to the reference results
in Table~\ref{tab:evaluation} on Sandy Bridge is due to the
application of a slightly different peak memory bandwidth.
It is striking that the ECM model is more optimistic than Roof\/line
for this benchmark, but this only shows that the latter is based
on a measured bandwidth value (with the very same microbenchmark kernel);
on the other hand it is known that the ECM model is slightly
optimistic with very tight, data-bound kernels, presumably because
of hardware prefetchers not keeping up with the in-core execution
in this case. Since the model assumes perfect prefetching, this
deviation could be fixed by adding latency penalty cycles to the
single-core execution.

\begin{lstlisting}[
    caption={Kernel code for the Sch\"onauer Triad},
    label={lst:triad_code},]
double a[N], b[N], c[N], d[N], s;

for(int i=0; i<N; ++i)
    a[i] = b[i] + c[i] * d[i];
\end{lstlisting}

\section{Related Work}\label{sec:related_work}

There are many performance modeling tools that rely on hardware
metrics, statistical methods, curve fitting, and machine learning, but
there are currently only four projects in the area of automatic
analytic modeling that we know of: PBound, ExaSAT, Roof\/line Model
Toolkit and MAQAO.

Narayanan et al.\ \cite{pbounds2010} describe a tool (PBound) for
automatically extracting relevant information about execution
resources (arithmetic operations, loads and stores) from source
code. They do not, however, consider cache effects and parallel
execution, and their machine model is rather idealized.  Unat et
al.\ \cite{Unat01052015} introduce the ExaSAT tool, which uses
compiler technology for source code analysis. They also employ ``layer
conditions'' \cite{stengel14} to assess the real data traffic for
every memory hierarchy level based on cache and problem sizes. They
use an abstract simplified machine model, whereas our \kerncraft\ tool
employs Intel IACA (see Sect.~\ref{sec:incore}) to generate more
accurate in-core predictions. On the one hand this
restricts \kerncraft\ in-core predictions to Intel CPUs, but on the
other hand provides predictions from the actual machine code
containing all the optimizations provided by the
compiler. Furthermore, ExaSAT is restricted to the Roof\/line model
for performance prediction.  On the other hand, being compiler-based,
ExaSAT supports full-application modeling and code optimizations,
which is out of scope in the current version of \kerncraft. It can
also incorporate communication (i.e., message passing) overhead.  Lo
et al.~\cite{rmt15} have recently introduced the ``Roof\/line Model
Toolkit,'' which aims at automatically generating hardware
descriptions for Roof\/line analysis. They do not support automatic
topology detection, however, and their use of compiler-generated loops
introduces an element of uncertainty.  Djoudi et
al.~\cite{djoudi2005maqao} started the MAQAO Project in 2005, which
uses static analysis to predict in-core execution time and combines it
with dynamic analysis to assess the overall code quality. It was
originally developed for the Itanium~2 but has since been adapted for
recent Intel64 architectures and the Xeon Phi. As with Kerncraft, MAQAO
currently supports only Intel architectures. The memory access analysis is based on
dynamic run-time data, i.e., it requires the code to be run on
the target architecture.

There is to our knowledge no tool comparable to IACA for non-Intel
CPUs. Fallback support for in-core predictions based on automatic code
analysis and architectural information (as done, e.g.,
in \cite{Unat01052015}) is work in progress.

\section{Summary and Outlook}\label{sec:summary}

The power and utility of analytic performance modeling is undisputed,
but the construction of accurate models is a tedious task that
requires considerable time and experience.  The \kerncraft\ tool
enables analytic (Roof\/line or ECM) modeling of loop kernels on
modern CPU architectures under well-defined conditions.  We have shown
that predictions generated by \kerncraft\ concur with published
numbers on a variety of codes and two modern Intel architectures.
However, it must be emphasized that \kerncraft\ is not a ``black-box''
tool that can be applied blindly without background knowledge.
In contrast, it is the deviation of generated models from experience
or from measured results that sparks new insight. This is impossible
without a basic understanding about the software and the capabilities
of the hardware it is running on.

Currently the applicability of the tool is restricted by dependencies
on the Intel-specific IACA tool and compilers.
But even without those \kerncraft\ can
still predict data locality and cache behavior.
Since the basic principles of the modeling process implemented
in \kerncraft\ are widely applicable, support for other architectures
is a matter of extending the ECM model appropriately.
Currently we are working on adapting the model to ARM-based chips.
To further enhance the tool in the short term we will support
single precision floating-point numbers,  passing pragma directives to the
compiler, and allowing for other compilers than ICC.
On the ECM model side, non-inclusive cache hierarchies and non-temporal stores
need further evaluation but will be implemented as well. It will be 
a more difficult task to find or build a suitable replacement for IACA
and to predict the data transfer properties of non-contiguous access patterns.
The simplistic cache model in \kerncraft\ may be extended 
to non-LRU cache replacement policies and non-fully associative
caches, but we consider such additions less crucial.

To combine dynamic analysis with performance modeling, we plan to
introduce phenomenological models where data traffic and bandwidth
measurements are used to improve the validation procedure and to
capture effects that can not be easily modeled, such as
irregular memory accesses as found, e.g., in sparse-matrix
algorithms. The required infrastructure to take the necessary data is
already functional in the current version of \kerncraft.

\section*{Acknowledgments}

Discussions with Johannes Hofmann are gratefully acknowledged.

\pagebreak
\bibliographystyle{abbrv}
\bibliography{paper}

\end{document}